# Tailoring the crack-tip microstructure: A novel approach

Khilesh Kumar Bhandari [a,*], Arka Mandal [a], Md. Basiruddin Sk [a], Arghya Deb [b], Debalay Chakrabarti [a]

[a] Department of Metallurgical and Materials Engineering, Indian Institute of Technology, Kharagpur, Kharagpur-721302, India

[b] Department of Civil Engineering, Indian Institute of Technology, Kharagpur, Kharagpur-721302, India

**\* Corresponding author (Last name, First name):**

**Bhandari, Khilesh Kumar**
Email id: khileshbha009@gmail.com, khileshkrbhandari@iitkgp.ac.in
Address: Department of Metallurgical and Materials Engineering, IIT Kharagpur, Kharagpur – 721 302, West Bengal, India.
Contact details: (+91) 9547976523

**Co-authors (Last name, First name):**

**Mandal, Arka**
Email id: arkametbesu@gmail.com

**Sk, Md. Basiruddin**
Email id: basiruddin.sk@gmail.com

**Deb, Arghya**
Email id: arghya@civil.iitkgp.ac.in

**Chakrabarti, Debalay**
Email id: debalay@metal.iitkgp.ernet.in

# Tailoring the crack-tip microstructure: A novel approach

Khilesh Kumar Bhandari [a,*], Arka Mandal [a], Md. Basiruddin Sk [a], Arghya Deb [b], Debalay Chakrabarti [a]

[a] Department of Metallurgical and Materials Engineering, Indian Institute of Technology, Kharagpur, Kharagpur-721302, India

[b] Department of Civil Engineering, Indian Institute of Technology, Kharagpur, Kharagpur-721302, India

## Abstract

This investigation demonstrates a way to innovatively modify the ferritic microstructure at a local scale, particularly at the failure prone area such as Charpy V-notch (CVN) root. Tensile pre-strain (PS) up to 6 percent and 12 percent were employed before annealing (An) the samples at 650°C for 15 minutes. Ferrite grain size increased sharply and gradually (along the distance ahead of the notch root) within the microstructurally modified region in 6-12 percent pre-strained and annealed samples, respectively. Critical strain which promotes strain induced boundary migration (SIBM), was found to be 0.1 which resulted in abnormally coarse ferrite grains.

## Introduction

Failure of a structural component takes place when a crack propagates under application of an external load. The brittle failure is more catastrophic in nature than the ductile failure. Since, brittle failure occurs via propagation of nucleated micro-cracks [1], the role of microstructural parameters needs to be carefully reviewed. So far, the studies [2–6] mostly investigated the effect of several microstructural parameters, such as grain size, second-phase particle size, fraction of different microstructural constituents and crystallographic orientation on the resistance to brittle (cleavage) crack propagation. An overall fine grain or bimodal grain structure is reported to possess better mechanical properties in terms of the strength-toughness combination [7,8]. In a bimodal grain structure, the fine grains provide high strength and resistance to crack propagation, while the coarse grains contribute to strain hardening ability and the overall ductility [9,10]. Nevertheless, producing such a functionally graded microstructure over a large structural section/component is a huge challenge.

Several studies have been carried out to understand the effect of pre-strain on various mechanical properties and the associated failure mechanisms [11–17]. Pre-straining at room temperature can alter dislocation substructure by acute dislocation interaction and multiplication [11], resulting in an increased yield and tensile strength [12,13]. However, ductility presumably follows an inverse relationship with the degree of pre-straining [12–14]. Fracture toughness has been reported to be decreased continuously as a function of pre-straining [16,17], while some investigators also found that it decreased after a certain small pre-strain level [15]. Besides, pre-straining makes the material susceptible to brittleness and material can fail at a smaller strain on subsequent reloading, as reported for low carbon steel [14].

Unlike above mentioned studies where homogeneous pre-straining is practised, loading/pre-straining of notched/pre-cracked specimen inevitably influences the stress and strain distribution by forming plastic zone ahead of crack tip [18]. The concept of stress singularity as estimated by elastic stress field equations is not effective at crack tip, due to the presence of the plastic zone [19]. By using the elastic stress field equation for $\sigma_{yy}$ in plane $\theta = 0$, the crack tip plastic zone size ($r_p^*$) can be quantitatively estimated by equation: $r_p^* = K_I^2 / 2\pi\sigma_{ys}^2$, where $K_I$ is the stress intensity factor in crack opening mode and $\sigma_{ys}$ is the yield stress. Later, the crack was considered to be longer than its actual size (i.e. $a_{eff} = a_{act} + \delta$), by Irwin, in order to eliminate the ambiguity in plastic zone size estimation and it was found to be twice of the $r_p^*$ [20]. However, the actual estimation of the plastic zone shape can be obtained by considering the entire range of theta $\theta$ ($\pm 90°$), instead of a circular shape.

When a material deforms, inhomogeneity is induced in the form of dislocations and point defects. To bring the material in its lower energy configuration, subsequent annealing results in partial recovery by the annihilation and rearrangement of those defects. Recovery is a series of events where the initially tangled dislocations first lead to cell formation [21], then sub-grain development and its growth. These phenomena are strongly dependent on factors like amount of prior strain, deformation and annealing temperature. In addition, recrystallization is another energy minimization process (governed by internal stored energy in cold worked metal) where new strain-free grains form and grow subsequently.

Recrystallization is discontinuous and generally involves large strain relaxation as compared to recovery. The recrystallization kinetics is very similar to that of phase transformation [21] and can get affected by processing parameters (amount of prior strain, strain rate, deformation temperature, mode of deformation, annealing time and temperature) and by microstructural parameters (initial grain size, second-phase particle, grain orientation) [22–24].

Recrystallized grains having high surface energy, which makes the structure unstable, can further grow. Grain growth decreases the grain boundary area, resulting in reduction of the total surface energy. This process can be classified either as normal or abnormal grain growth [25]. Former is a continuous process where all grains constituting the material grow uniformly. Latter occurs when normal grain growth is inhibited, except for some grains having favourable condition to grow at the expense of neighbouring recrystallized grains [26]. Abnormal grain growth is more pronounced in alloys having second-phase particles, causing Zener pinning [27–29]. Other factors, e.g. crystallographic texture and surface effect can also contribute to this process [26,30]. Abnormal grain growth can also happen in a partially recrystallized microstructure, where the boundary of the strain-free recrystallized grain moves to consume the neighbouring strain matrix containing deformed/recovered grains. This phenomenon is known as SIBM [31,32].

Considering the difficulty to generate the bimodal grain structure in large component, the present study aims to develop suitable microstructure at a local scale, particularly at the failure prone area (such as the crack front) in a single phase ferritic steel by tensile pre-straining and subsequent annealing.

## Materials and Methods

Standard sub-size Charpy V-notched (5x10x55 mm) specimens were extracted from a fully ferritic steel with the average composition of 0.03 C, 0.25 Mn, 0.01 S, 0.02 P, 0.008 Si, 0.07 Al, 0.02 Cr, 0.02 Ni, and 0.0036 Ni (wt. %). Investigated material consisted of equiaxed ferrite grains with average grain size of 37 ± 7 µm. Tensile pre-strain up to 6 percent (6%PS) and 12% (12%PS) were employed before annealing (An) at 650°C for 15 minutes under inert atmosphere (to prevent any damage at the notch tip). A sample without pre-straining (0%PS) was also annealed at the same condition for comparison. An elastic-plastic finite element (FE) simulation was performed in ABAQUS/explicit software to quantitatively estimate the stress-strain distribution ahead of notch root with root radius of 0.25 mm after pre-straining. Microstructural characterization was performed using optical microscopy, electron back-scatted diffraction (EBSD) and nanoindentation after conventional preparation of the samples.

**Results**

Basic power hardening law, $\sigma = K\varepsilon;\ \sigma < \sigma_{ys}\ and\ \sigma = K\varepsilon^n;\ \sigma \geq \sigma_{ys}$, was considered for FE simulation in ABAQUS, where, elastic modulus, E = 210 GPa and Poisson's ration, ν = 0.27. Plastic strain contour maps revealing the plastic zone in **figure 1(a, b)** display that the farthest boundary from notch tip is located at an angle of ~58.5° and ~48° in 6%PS and 12%PS samples, respectively, in comparison to 69°, as reported by Tuba [33]. The increase in pre-strain significantly increases the plastic zone size as in **figure 1(a, b)**. The shape of the plastic zone obtained from FE simulation in this investigation is in good agreement with the findings reported by Tuba [33] and Hahn and Rosenfield [34]. **Figure 1(c, d)** shows the stress and strain distribution as a function of distance from the notch tip in 6 percent and 12 percent pre-strained samples. An increase in stress and strain values is observed as the degree of pre-strain increased, while the distribution followed same pattern. The maximum normal stress, $\sigma_{yy}$, for both pre-strained samples has values more than 2.5 times the general yield stress, **figure 1(c)**, which is close to the theoretical values of 3 [19]. On the other hand, the plastic strain, $\varepsilon_{yy}$, attains its maxima at the notch tip and decreases sharply moving away from the notch tip (**figure 1(d)**).

Ferrite grain size remained unchanged, however, a minimal directionality in grain morphology just ahead of the notch tip was observed in the pre-strained samples. Annealing of 0%PS sample did not show considerable microstructural alteration in **figure 2(a)**. Moreover, **Figure 2(b, c)** clearly display the regions over which microstructural modification have occurred after annealing in accordance with the plastic zone size of the 6-12% pre-strained sample. The modified region revealed substantial gradient in grain size; from finer ferrite grains (~25 µm) at notch root to abnormally coarse grains (~200-300 µm) at the end of the modified region. Interestingly, 12%PS-An sample exhibited a gradual increase in grain size with the coarsest grains (~200-250 µm) located towards the end of the modified region (3 mm away from notch root), **figure 2(c)**. In contrast, 6%PS-An sample displayed steep grain-size gradient with the coarsest grains (~270-300 µm) situated at a distance of just 1 mm away from the notch tip, **figure 2(b)** (refer grain size distribution (right) in **figure 2(b, c)**).

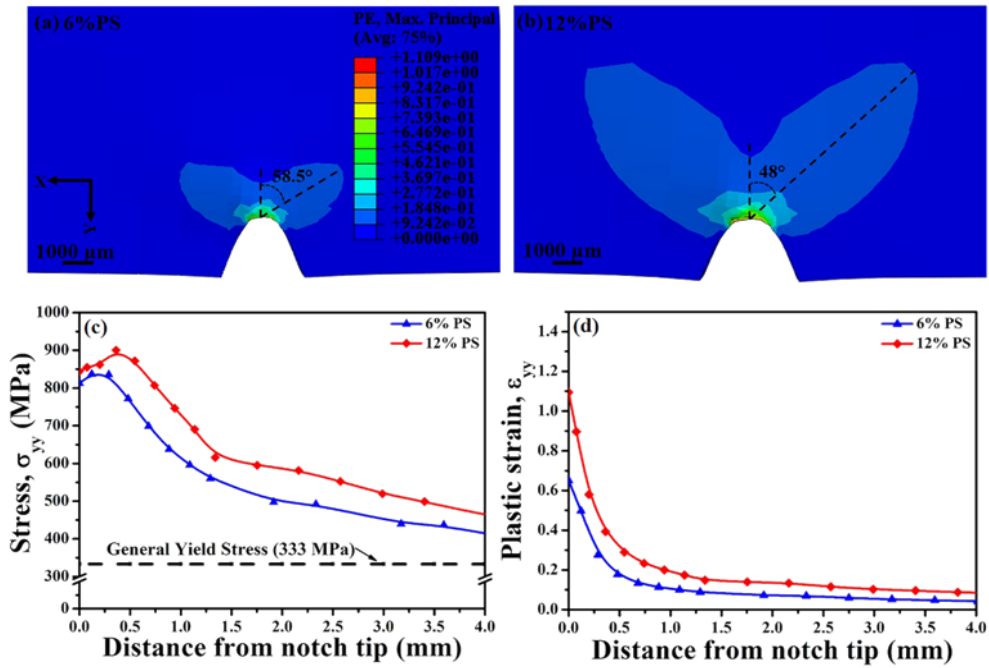

**Figure 1.** Plastic strain ($\varepsilon_{max}$) contour showing notch tip plastic zone, obtained from FE simulation using ABAQUS/explicit for: (a) 6% pre-strained (6%PS), and (b) 12% pre-strained (12%PS) samples. Stress ($\sigma_{yy}$) distribution (c), and strain ($\varepsilon_{yy}$) distribution (d) ahead of notch root after pre-straining.

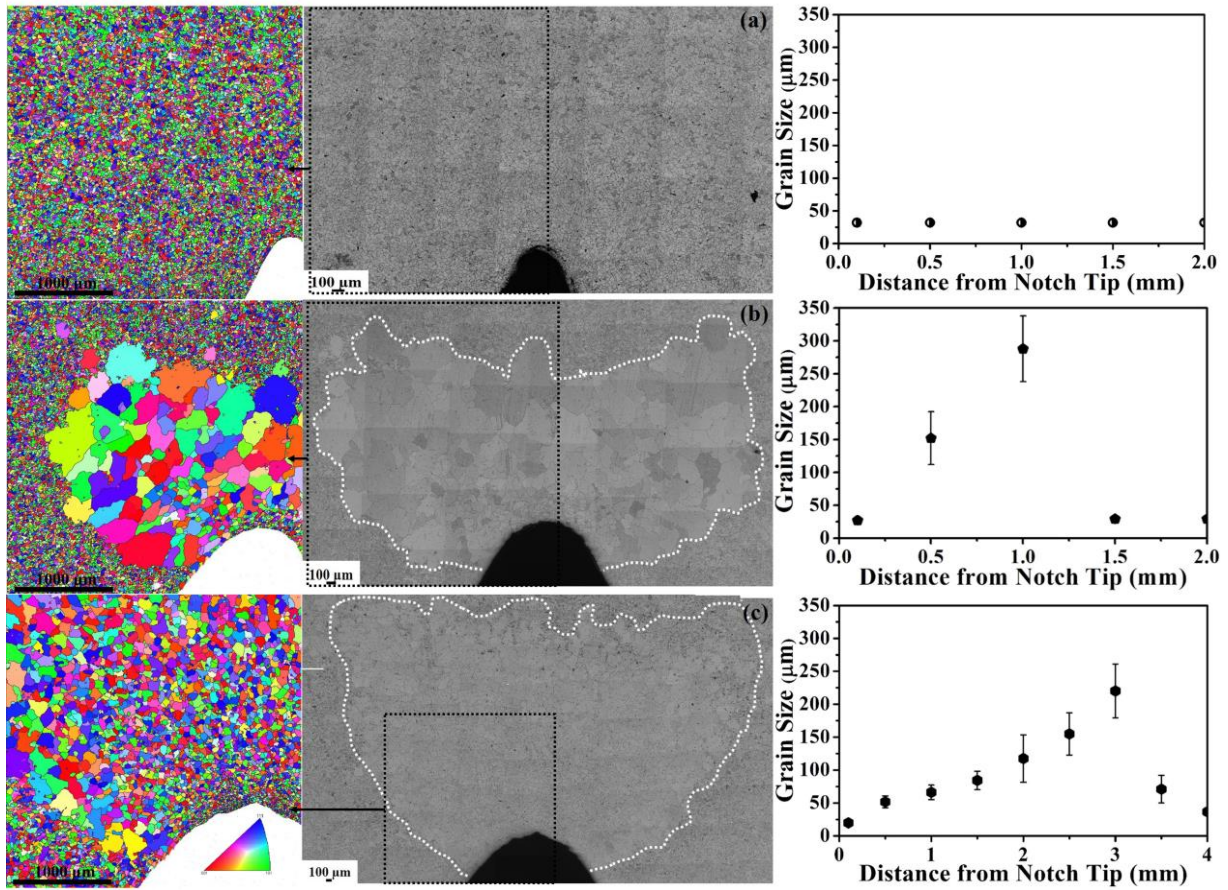

**Figure 2.** IPF ND maps revealing enclosed modified region (left), optical microstructure showing modified ferrite grains (centre), and corresponding ferrite grain size distribution with distance from notch tip (right) in pre-strained and annealed (PS-An) samples: (a) 0%PS-An, (b) 6%PS-An, and (c) 12%PS-An.

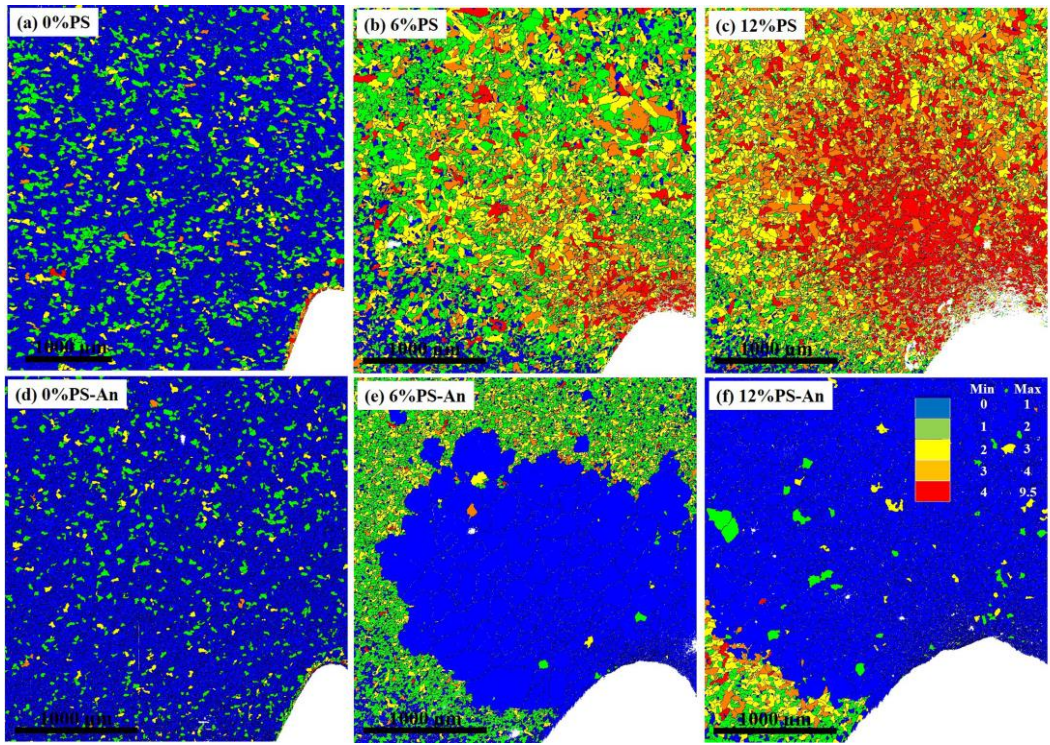

**Figure 3**. Grain orientation spread (GOS) maps, obtained by EBSD analysis to qualitatively display strains in the investigated samples: (a) 0%PS, (b) 6%PS, (c) 12%PS, (d) 0%PS-An, (e) 6%PS-An, and (f) 12%PS-An.

Grain orientation spread (GOS) map in **figure 3** is an effective measure of the strain in terms of misorientation, particularly within a grain. Average GOS value is calculated by averaging deviation between the orientation of each data point within a grain and average orientation of that grain. A small GOS value (0.56) represents the characteristic of an annealed structure in 0%PS sample, **figure 3(a)**. The pre-straining of CVN samples induced plastic deformation of material ahead of notch root, forming plastic zone and resulted in higher GOS value in 6%PS (2.26) and 12%PS (2.80) samples, **figure 3(b, c)**, ahead of the notch tip. Supposedly, GOS value did not change in 0%PS-An sample, **figure 3(d)**, since there was no prior strain which could be released during annealing. The modified region obtained after annealing the 6%PS and 12%PS samples, **figure 3(e, f)**, divulged strain-free grains as a result of strain relaxation. Additionally, the unmodified region surrounding this modified region still possessed deformed structure as revealed in the corresponding GOS maps.

Average in-grain misorientation (avg. GOS distribution) ahead of the notch root is calculated by creating square subsets at distances and plotted as a function of distance from the notch tip, **figure 4(a)**. GOS values follow an increasing trend with distance in 6%PS-An and 12%PS-An samples, starting from the lowest (0.28) at the notch tip. Presence of a deformed region ahead of the modified region, which remained unaltered after annealing, was confirmed by high GOS values (1.4 in 6%PS-An and 1.8 in 12%PS-An). On the contrary, pre-strained CVN samples (6%PS and 12%PS) demonstrate an opposite i.e. decreasing trend in their GOS values in **figure 4(a)**, since strain accumulation is the maximum at the notch root and decreases with distance. A uniform distribution of GOS value as the function of distance in as-received (0%PS) and subsequent annealed (0%PS-An) samples is also shown for comparison.

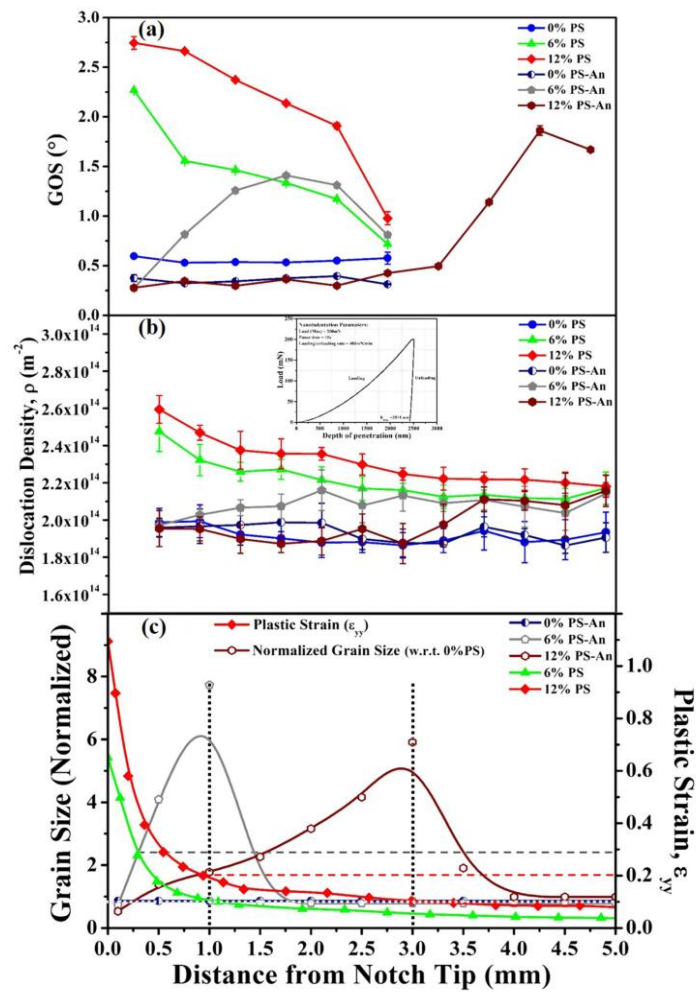

**Figure 4**. (a) Average GOS distribution, and (b) Dislocation density distribution ahead of notch root. (c) Plot predicting critical strain for strain-induced boundary migration (SIBM) using grain size distribution (normalised w.r.t. 0%PS grain size) and notch tip strain distribution curve.

Dislocation density (DD) obtained via nanoindentation technique following Nix and Gao model [35,36] confirms the presence of highly deformed grains (DD ~$2.48 \times 10^{14} \pm 1.1 \times 10^{13}$) just ahead the notch tip in pre-strained samples, **figure 4(b)**. DD decreases on moving away from the notch tip, as expected. An opposite, increasing trend with the increase in distance from notch tip is observed in the annealed specimens having lowest DD ($\sim 1.95*10^{14} \pm 3.1\times10^{12}$) at the notch tip, **figure 4(b)**.

## Discussion

The present work is performed in single phase ferritic steels just to curtail microstructural variables other than the grain size. Pre-straining promoted plastic flow resulting in increase in the grain aspect ratio from 1.1 to 1.3. Subsequent annealing of the deformed structure releases the accumulated strain by mechanisms like recovery, recrystallization and grain growth. It is well recognized that the restoration mechanism such as recrystallization and SIBM depends primarily on amount of prior deformation, subsequent annealing temperature and time. Since, the annealing temperature and time was kept constant, prior strain was the sole parameter controlling the final ferrite grain size and the grain size distribution. Presence of fine, equiaxed and strain free ferrite grains ahead of the notch root in figure 2(b, c) is due to complete recrystallization which occurs when the local strain exceeds a critical strain. However, strain below this critical value could not yield complete recrystallization and ended up in structural restoration by grain coarsening following SIBM toward the end of the modified region over a distance of ~ 1.20 mm, away from the notch root, figure 2(b, c). Abnormal grain growth by SIBM was reported earlier in several studies [31,37,38], where locally accumulated strain could not reach the critical strain (i.e. driving force) required for complete recrystallization.

GOS distribution plot shown in figure 4(a) confirmed complete recrystallization at the notch tip and a deformed region just ahead of the modified region in the annealed samples. Although, the mechanistic calculations clearly define the plastic zone size, its boundary may not necessarily be a sharp one, rather a diffused boundary spread over a distance. The trends found in GOS distribution plots are in good agreement with the dislocation density profile as a function of distance ahead of a crack-tip for all the samples, figure 4(a, b). Therefore, GOS maps can effectively be used for qualitative estimation of the local strain in both pre-strained and annealed condition.

Since, the critical strain for recrystallization cannot be estimated solely by optical micrographs, an attempt is made in order to find out the critical strain for SIBM by analysing modified ferrite grain size (of annealed samples) and plastic strain (of pre-strained samples) distribution plots together, figure 4(c). It shows the variation in ferrite grain size (normalised with respect to the grain size of the as-received sample) as a function of plastic strain ahead of the notch tip within the modified region. The plastic strain corresponding to the coarsest ferrite grains (at a distance of 1 mm and 3 mm in 6%PS-An and 12%PS-An samples, respectively, from the notch tip) is found to be 0.1 in both 6%PS and 12%PS samples. The presence of these coarse grains in annealed samples is a consequence of SIBM which occurs at a strain lower than that necessitated for complete recrystallization. With this it can be concluded that there exists a critical strain, 0.1 in this investigation, which promotes SIBM when annealed. The microstructural modifications extended up to a distance decided by the local strain level (down to $\varepsilon = 0.1$), beyond which the local strain is insufficient (but not necessarily zero) to drive any microstructural restoration. As the local strain exceeds 0.1, complete recrystallization and associated grain refinement governed the microstructural restoration.

The positive effect of functionally graded microstructure (finer or bimodal type) on various properties are stated in many studies before [5–8]. Release of residual strain and grain structure modification are expected to influence the fracture toughness positively by resisting the crack propagation through crack blunting. Future investigation aims to look into the effect of pre-straining and subsequent annealing on impact transition behaviour and other mechanical properties. Devising a model to predict critical strain with ferrite grain size as an input parameter can be another future aspect.

## Conclusions

In summary, this investigation unveils a new methodology of tailoring microstructure at a local scale, especially at failure prone area via pre-straining and subsequent annealing, since developing a suitable microstructure over a large structural component is challenging. A method to predict critical strain, using grain size (after annealing) and plastic strain (after FE simulation) distribution, promoting SIBM is also presented in this work. The critical strain for SIBM has been found out to be 0.1 and it decreased gradually hereafter, forming a diffused boundary of the plastic zone. It is noteworthy to mention that if a pre-existing crack can be detected within a structural component by non-destructive testing, suitable heat treatment can be employed following the current approach to modify the local microstructure and enhance the life of the whole component.

## Acknowledgements:

Authors acknowledge support from Dr B. Syed, TATA steel, Jamshedpur for providing material for this investigation. Authors gratefully concede the thoughtful discussion with late Prof. John F. Knott, Birmingham, UK and Dr. S. V. Kamat, DS & DG – NS & M, Defence Metallurgical Research Laboratory (DMRL) Hyderabad, India.